\documentclass{article}

\usepackage[T1]{fontenc}
\usepackage{microtype}
\usepackage{graphicx}
\usepackage{subcaption}
\usepackage{booktabs}
\PassOptionsToPackage{hyphens}{url}
\usepackage[breaklinks=true]{hyperref}

\usepackage[accepted]{icml2026}

\makeatletter
\renewcommand{\Notice@String}{Accepted at the 6th Muslims in ML (MusIML) Workshop at ICML 2026.}

\renewcommand{\printAffiliationsAndNotice}[1]{\global\icml@noticeprintedtrue%
  \stepcounter{@affiliationcounter}%
  {\let\thefootnote\relax\footnotetext{\hspace*{-\footnotesep}\ificmlshowauthors #1\fi%
      \forloop{@affilnum}{1}{\value{@affilnum} < \value{@affiliationcounter}}{
        \textsuperscript{\arabic{@affilnum}}\ifcsname @affilname\the@affilnum\endcsname%
          \csname @affilname\the@affilnum\endcsname%
        \else
          {\bf AUTHORERR: Missing \textbackslash{}icmlaffiliation.}
        \fi
      }.%
      \ \\
      \Notice@String
    }
  }
}
\makeatother

\usepackage{amsmath}
\usepackage{amssymb}
\usepackage{mathtools}
\usepackage{amsthm}

\usepackage[capitalize,noabbrev]{cleveref}

\usepackage[most]{tcolorbox}
\tcbuselibrary{breakable, skins}
\usepackage[table]{xcolor}
\usepackage{float}
\definecolor{promptframe}{RGB}{36,80,143}
\definecolor{promptback}{RGB}{246,249,253}

\newtcolorbox{promptbox}[1]{
  enhanced jigsaw, breakable,
  colback=promptback,
  colframe=promptframe,
  coltitle=white,
  colbacktitle=promptframe,
  title={\strut #1},
  fonttitle=\bfseries\sffamily\small,
  fontupper=\small,
  arc=0.6mm,
  boxrule=0.4pt,
  titlerule=0pt,
  left=3mm, right=3mm, top=2mm, bottom=2mm,
  parbox=false,
}

\theoremstyle{plain}

\theoremstyle{definition}

\theoremstyle{remark}

\icmltitlerunning{CRC-Screen: Certified DNA-Synthesis Hazard Screening Under Taxonomic Shift}

\begin{document}

\twocolumn[
  \icmltitle{CRC-Screen: Certified DNA-Synthesis Hazard Screening \\ Under Taxonomic Shift}

  \icmlsetsymbol{equal}{*}

  \begin{icmlauthorlist}
    \icmlauthor{Najmul Hasan}{uncp}
  \end{icmlauthorlist}

  \icmlaffiliation{uncp}{University of North Carolina at Pembroke}

  \icmlkeywords{conformal prediction, biosecurity, DNA synthesis screening, calibrated classification}

  \vskip 0.3in
]

\printAffiliationsAndNotice{}

\begin{abstract}
  DNA-synthesis providers screen incoming orders by searching the
  requested sequence against curated hazard lists. We show that this
  baseline collapses to a $100\%$ false-flag rate when the hazardous
  sequence comes from a taxonomic family absent from the reference
  set: under Conformal Risk Control's certified miss-rate constraint,
  a low-discrimination signal forces the threshold below the entire
  test-benign mass. We compose three signals derived from a synthesis
  order's public annotation: $k$-mer Jaccard similarity to known
  toxins, the trimmed-mean score of a five-LLM judge panel, and cosine
  similarity to clustered embedding centroids. Fused under a monotone
  logistic aggregator and calibrated by Conformal Risk Control, the
    resulting screener certifies $\mathbb{E}[\mathrm{FNR}] \le \alpha +
  \mathrm{TV}$, where the additive term is the calibration-to-test
  distribution shift under family holdout (a certified ceiling of
  $24$--$49\%$ across folds). Across ten
  leave-one-taxonomic-family-out folds at $\alpha=0.05$ on UniProt
  KW-0800 reviewed toxins, the calibrated screener achieves $0\%$
  \emph{empirical} test miss rate on every fold and $0\%$ test
  false-flag rate on nine of ten folds. The bound's finite-sample slack $1/(n_\mathrm{cal}+1)$
  caps the certifiable miss rate at $1.77\%$ on our $200$-hazard
  subsample; reaching procurement-grade $\alpha=10^{-3}$ requires an
  $18\!\times$ larger calibration set, which the full reviewed UniProt
  KW-0800 corpus is large enough to deliver. The binding constraint
  on certifiable DNA-synthesis screening is calibration data, not
  algorithms.\ \textit{Code:} \url{https://github.com/najmulhasan-code/crc-screen}
\end{abstract}

\begin{figure}[!t]
  \centering
  \includegraphics[width=\columnwidth]{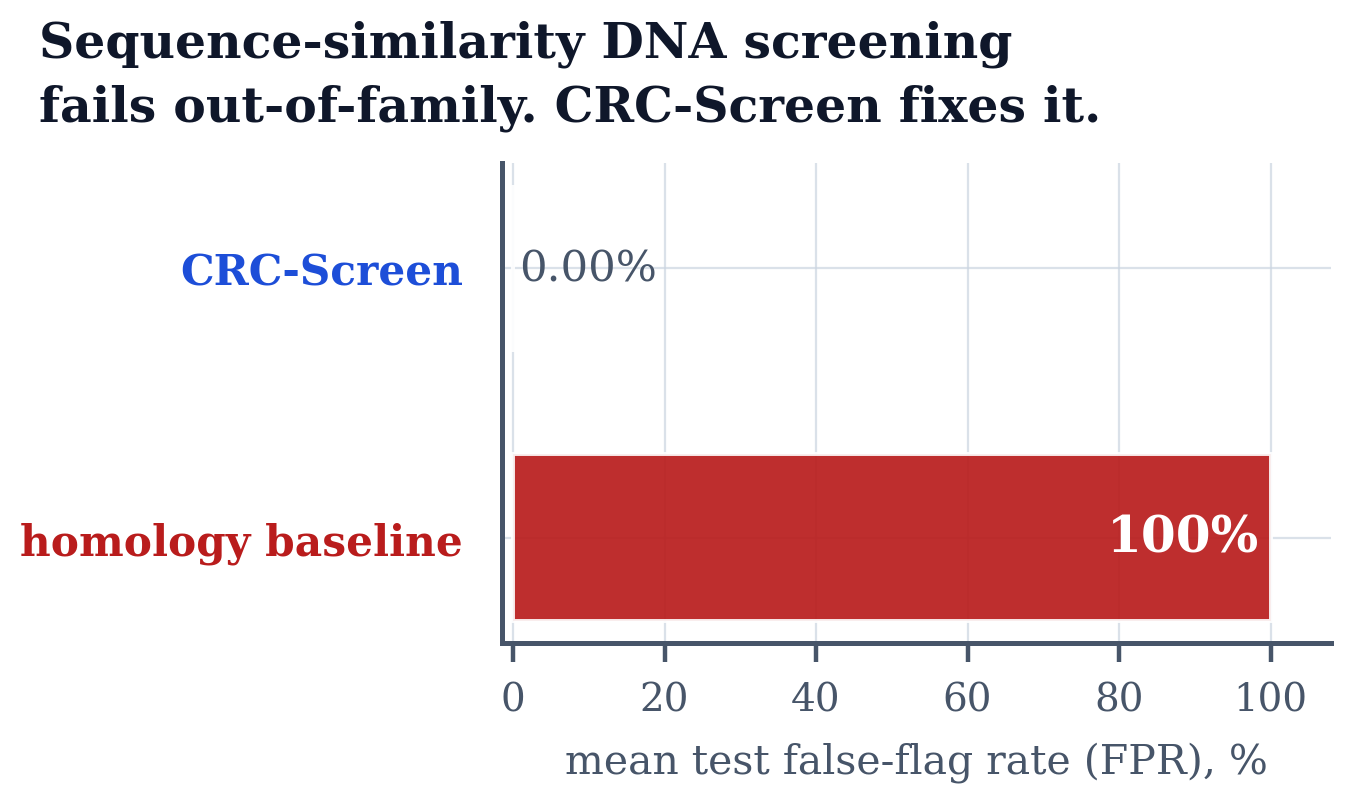}
  \caption{
    Sequence-similarity-only screening flags every benign in
    out-of-family folds ($100\%$ FPR); CRC-Screen drops this to
    $0\%$ while certifying $\mathbb{E}[\mathrm{FNR}]\le\alpha+\mathrm{TV}$ at
    $\alpha=0.05$, mean across ten leave-one-taxonomic-family-out folds.
  }
  \label{fig:teaser}
\end{figure}

\section{Introduction}
\label{sec:intro}

DNA-synthesis providers are the last enforcement point before a
hazardous protein is built: an order for such a protein can in
principle be intercepted between the customer's design and the
synthesised molecule. The standard implementation of that bottleneck is
a sequence-similarity search against curated hazard lists derived from
regulatory inventories and toxin databases. This baseline was built for
a threat model in which the hazardous order looks, at the level of
amino-acid sequence, like a hazardous protein the screener has already
seen. Two trends now stretch that assumption. Generative models for protein
design \cite{madani2023progen,lin2023evolutionary}
produce variants that retain function while drifting in primary
sequence, and red-team studies have begun to evaluate whether
language-model assistance offers operational uplift to non-state
actors planning biological attacks, finding no significant uplift
from current models but flagging trajectory risk for future systems
\cite{mouton2024operational}. Public toxin databases are fragmented
across specialised resources \cite{jungo2005toxprot,kaas2012conoserver},
with hazardous proteins from understudied taxonomic families sparsely
represented. Both
trends shift weight onto the out-of-family case, which is precisely
the case in which a sequence-similarity baseline weakens.

We make this concrete with a leave-one-taxonomic-family-out evaluation
on UniProt KW-0800 reviewed toxins. Held out one family at a time, the
sequence-similarity signal is too weak to separate hazards from benigns;
Conformal Risk Control, forced to certify a miss-rate ceiling, has no
choice but to push its threshold so low that every test benign is
flagged. The result is a flag-everything regime: $100\%$ test FPR on
every fold
(\cref{fig:teaser}, red bar).

The fix we study is composition. The same public annotation that
accompanies a synthesis order, comprising name, organism,
controlled-vocabulary keywords and a free-text function description,
makes two further signals available: a five-LLM panel reads the
annotation and returns a hazard probability, and the text-embedding
distance to clustered embeddings of known toxins gives a smooth proxy
for functional proximity. Composing the three signals under a monotone
logistic aggregator and then calibrating the decision threshold by
Conformal Risk Control \cite{angelopoulos2024conformal} restores
certified $\mathbb{E}[\mathrm{FNR}]\!\le\!\alpha+\mathrm{TV}$, and on our evaluation the calibrated screener achieves $0\%$ empirical test miss rate on every leave-one-family-out fold at $\alpha=0.05$, with $0\%$ test
false-flag rate on nine of ten folds and one flagged benign on
Actiniidae. A signal-by-signal ablation shows the LLM panel and the
embedding signal are jointly sufficient; adding sequence homology back
to that pair raises mean FPR by half a point with no recall gain.

The paper contributes four results. First, under taxonomic-family
holdout, $k$-mer-Jaccard sequence-similarity screening incurs a
$100\%$ false-flag rate at any non-trivial $\alpha$, an empirical
consequence of Conformal Risk Control's coverage requirement on a
low-discrimination signal that does not close under $\alpha$-tuning.
Second, a leak-controlled per-fold protocol (\cref{sec:method:crc})
prevents the held-out family's hazards from leaking into their own
scoring through the homology and embedding reference sets. Third, an
off-the-shelf composition of $k$-mer Jaccard, a five-LLM panel with
trimmed-mean aggregation, and an embedding-centroid distance, fused
by a monotone logistic aggregator and calibrated by Conformal Risk
Control, certifies $\mathbb{E}[\mathrm{FNR}]\!\le\!\alpha+\mathrm{TV}$ at $\alpha=0.05$ with $0\%$ empirical miss rate on every fold and $0\%$
empirical false-flag rate on nine of ten leave-one-taxonomic-family-out
folds. Fourth, the data budget that decides what $\alpha$ is reachable:
at $n_\mathrm{cal\,haz}\!\approx\!55$ the slack term
$1/(n_\mathrm{cal\,haz}+1)$ floors the certifiable $\alpha$ at
$1.77\%$, and procurement-grade $\alpha\!=\!10^{-3}$ requires
$n_\mathrm{cal\,haz}\!\ge\!999$, an $18\!\times$ gap that the full
reviewed UniProt KW-0800 corpus has the size to close.

\section{Conformal Risk Control and the screening status quo}
\label{sec:background}

\subsection{From coverage to risk}
\label{sec:bg:crc}

Conformal prediction
\cite{vovk2022algorithmic,angelopoulos2023gentle} converts any
black-box predictor into a procedure with finite-sample coverage
guarantees by calibrating a threshold on a held-out exchangeable
calibration set, with distribution-free regression and classification
specialisations now standard
\cite{lei2018distributionfree,romano2019conformalized,sadinle2019least,cauchois2021knowing}.
The classical guarantee is on miscoverage: the constructed prediction
set covers the truth with probability at least $1-\alpha$. Risk-control
extensions move the guarantee from a coverage event to a bounded loss
\cite{bates2021distributionfree}, and Conformal Risk Control
\cite{angelopoulos2024conformal} in particular generalises miscoverage
to any monotone, bounded loss. Given calibration losses
$L_1,\dots,L_n$ that are non-decreasing in a real-valued threshold
parameter $\tau$ and bounded above by $B$, the choice
$\widehat{\tau} = \sup\{\tau : \widehat{R}(\tau) + B/(n+1) \le \alpha\}$
satisfies $\mathbb{E}[L_{n+1}(\widehat{\tau})] \le \alpha$ on a fresh
exchangeable point, where $\widehat{R}$ is the empirical mean of the
calibration losses (\cref{eq:crc}). This is the standard non-increasing
form of CRC under the substitution $\lambda = -\tau$; we keep $\tau$
because the threshold is more natural for screening. For our screening
setting, $L_i$ is the false-negative indicator on hazard $i$, which is
non-decreasing in $\tau$ for any score $S$; we additionally constrain
the aggregator to be non-decreasing in each underlying signal
(\cref{sec:method:agg}) so that flag direction is consistent across
signals.

When the calibration and test distributions are not exchangeable the
guarantee picks up an additive correction. Theorem~2 of
\citet{barber2023conformal} bounds the deviation by a sum of weighted
total-variation terms over residual swaps, which generalises earlier
weighted-conformal results for covariate shift
\cite{tibshirani2019conformal}. We use a histogram TV between the
calibration and test score distributions as a coarse approximation of
that residual-swap quantity. The same calibration mindset underlies post-hoc score-rescaling for
classifier outputs \cite{platt1999probabilistic,guo2017calibration} and
selective classification with abstention thresholds
\cite{geifman2017selective}, which sit adjacent to our setting. Under leave-one-taxonomic-family-out
holdout we directly observe TV distances of $0.19$ to $0.44$ across
folds, so the bound's slack is dominated by this distribution-shift
term rather than by the finite-sample term $1/(n_\mathrm{cal\,haz}+1)$
at the alpha levels we test.

\subsection{What providers screen against}
\label{sec:bg:screening}

A DNA-synthesis order arrives at a provider as a sequence specification
plus customer metadata. Before fulfilling the order, providers in the
International Gene Synthesis Consortium (IGSC) and equivalents run a
sequence-similarity search against curated lists of pathogen and toxin
sequences, escalate flagged orders to human review, and sometimes
require additional customer attestation \cite{carter2015dna,diggans2019nextsteps}.
The technical core of this screening is the same alignment search
machinery used throughout computational biology, descended from BLAST
\cite{altschul1990basic,altschul1997gapped,camacho2009blastplus} and
its modern protein-scale successors such as DIAMOND
\cite{buchfink2021sensitive}, MMseqs2 \cite{steinegger2017mmseqs2},
USEARCH \cite{edgar2010search} and profile-HMM tools
\cite{eddy2011accelerated} indexed against family databases such as
Pfam \cite{mistry2021pfam}. The policy and biosecurity literature
documents two structural problems with this baseline: short fragments below
the alignment-search sensitivity floor escape detection
\cite{diggans2019nextsteps}, and
AI-designed protein variants that depart in primary sequence from
training distributions can fall below the same thresholds while
preserving function \cite{wittmann2025strengthening}.
 Our system retains the homology signal as one of
three inputs (\cref{sec:method:signals}) but does not rely on it for
discrimination under taxonomic-family holdout.

\section{Three signals, monotone fusion, calibrated threshold}
\label{sec:method}

\begin{figure*}[!t]
  \centering
  \includegraphics[width=\textwidth]{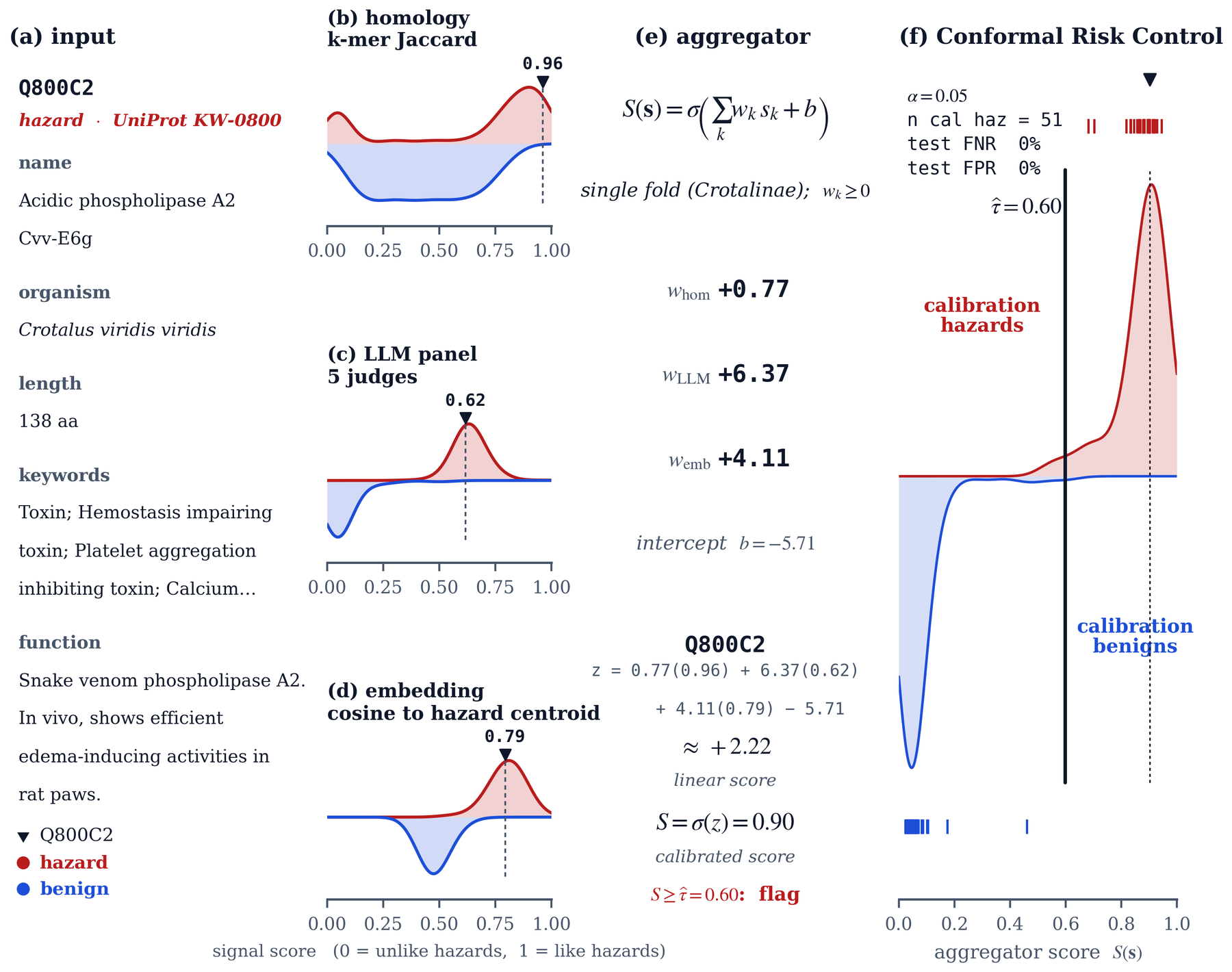}
  \caption{
    CRC-Screen takes a UniProt annotation through three signals, fuses
    them with a monotone logistic aggregator, and flags the order if
    the calibrated score $S$ exceeds the Conformal Risk Control
    threshold $\widehat{\tau}$; on the held-out \emph{Crotalinae}
    fold, this correctly flags Q800C2 with test FNR$=0\%$ and test
    FPR$=0\%$.
    (a) Input UniProt record (accession Q800C2, an acidic
    phospholipase A$_2$ from \emph{Crotalus viridis viridis}, labelled
    hazard via KW-0800).
    (b) to (d) Per-fold signal distributions across the $n{=}600$
    corpus, with the example's value marked by an inverted triangle.
    (e) The aggregator applied to Q800C2: the substituted weights and
    signal values yield $z\!\approx\!+2.22$, so $S=\sigma(z)=0.90$.
    (f) The CRC threshold chosen on the calibration densities:
    $\widehat{\tau}=0.60$ with $n_\mathrm{cal\,haz}=51$. Since
    $S\!\ge\!\widehat{\tau}$, Q800C2 is flagged.
  }
  \label{fig:method}
\end{figure*}

A synthesis order arrives with a public UniProt annotation: an accession,
the protein's name, the source organism, a controlled-vocabulary keyword
list, and a free-text function description. From that annotation we derive
three signals, fuse them with a monotone logistic aggregator, and pick the
flag-versus-pass threshold by Conformal Risk Control. The whole pipeline,
illustrated for one held-out family, is shown in
\cref{fig:method}.

\subsection{What each signal captures}
\label{sec:method:signals}

Each signal captures a different sense in which an order may resemble
a known hazard, and the resulting correlations are weak enough that a
linear aggregator gains from all three.

\paragraph{Homology signal $s_\mathrm{hom}$.} Sequence-similarity search
against curated hazard lists is the standard tool of current synthesis
screening, descended from BLAST \cite{altschul1990basic} and DIAMOND
\cite{buchfink2021sensitive}, with related index-based and
clustering-based variants \cite{steinegger2017mmseqs2,edgar2010search,suzek2015uniref}.
Our implementation is minimal: a $k$-mer Jaccard similarity
between the query sequence and each reference hazard, with $k\!=\!5$
amino acids; the per-query score is the maximum similarity over the
reference set, then rank-normalised across the $n{=}600$ corpus to a
value in $[0,1]$. Self-matches are excluded so that a hazard
under evaluation does not match itself. The choice of Jaccard over a
gapped alignment is conservative: it strips away the optimisations that
would let a sequence-similarity baseline look stronger than it is,
isolating the failure mode that motivates the rest of the system.

\paragraph{LLM panel score $s_\mathrm{LLM}$.} A panel of five large
language models reads the annotation text and returns a hazard
probability in $[0,1]$. The five models are Claude Opus~4.7
(Anthropic), GPT-5.4 (OpenAI), Llama 4 Maverick (Meta), Qwen 3.6 Plus
(Alibaba), and GLM 5.1 (ZAI), one per provider, so that systematic
refusals or scoring biases are unlikely to align across the panel. Each model receives the same
zero-shot prompt: a biosecurity-screening role, a four-level rubric tied
to standard regulatory categories, and a strict JSON output schema; the
prompt forbids generation of sequence data or synthesis instructions.
Each (sample, model) pair is queried $k\!=\!2$ times at temperature
$0.7$, and the per-model score is the median of the two runs (which
equals their mean at $k=2$). The panel score is the trimmed mean of
the five per-model scores, dropping the lowest and the highest and
averaging the three middle values. API failures, JSON-parse failures
and out-of-range scores are filled with the neutral value $0.5$ before
the trim, which absorbs at most one such fallback on each side.
The panel is an instance of the LLM-as-judge setup studied in
\citet{zheng2023judging} and developed for evaluation in
\citet{liu2023geval} and \citet{dubois2023alpacafarm}, differing in
its aggregation rule and application.

\paragraph{Embedding distance $s_\mathrm{emb}$.} Each annotation is
rendered as a labelled key--value string (name, organism, keywords,
function), passed through OpenAI's text-embedding-3-large model
in the lineage of contextual sentence and passage encoders
\cite{devlin2019bert,reimers2019sentencebert,karpukhin2020dense}, and
L\textsubscript{2}-normalised. Sequence-conditioned protein language
models \cite{rives2021biological,lin2023evolutionary,elnaggar2022prottrans}
offer a complementary representation; we use a text encoder over the
annotation rather than a sequence encoder over the protein because the
order's annotation arrives long before its sequence is committed to
synthesis. We then run $K$-means with
$K\!=\!\min(8,\ \lfloor n_\mathrm{train\,haz}/5\rfloor)$ on the embeddings
of the train-fold hazards alone, normalise the centroids, and define
$s_\mathrm{emb}$ as the maximum cosine similarity between the query's
embedding and any hazard centroid, clipped to $[0,1]$. Multiple
centroids accommodate the multi-modality of the hazard pool: toxins
from unrelated organisms occupy distant regions of the embedding space.

\subsection{Why the aggregator must be monotone}
\label{sec:method:agg}

Linear fusion of classifier outputs is standard
\cite{dietterich2000ensemble,caruana2004ensemble}.
We fuse the three signals with a logistic regression
\begin{equation}
  S(\mathbf{s}) \;=\; \sigma\!\Bigl(\,\sum_{k\in\{\mathrm{hom},\mathrm{LLM},\mathrm{emb}\}} w_k\,s_k \;+\; b\,\Bigr),
  \label{eq:agg}
\end{equation}
fit on the train-fold portion of each leave-one-family-out split, with
the constraint $w_k \!\geq\! 0$ for every signal. With only
non-negative weights the score $S$ is non-decreasing in every input
signal, so increasing any one of $\{s_\mathrm{hom}, s_\mathrm{LLM},
s_\mathrm{emb}\}$ can only raise the flag probability, not lower it; a
negative coefficient would invert that semantics for one signal and
break the agreement that composition is supposed to enforce.
Operationally, we fit an unconstrained logistic regression
\cite{pedregosa2011scikit}, drop the signal whose coefficient is most
negative, and refit on the remaining signals; this repeats until every
coefficient is non-negative, which always terminates because the empty
model trivially satisfies the constraint. The intercept $b$ is
unconstrained.

\subsection{Calibrating the threshold without leaking the test family}
\label{sec:method:crc}

Given calibration scores $\{S_i\}_{i\in\mathrm{cal}}$ and labels
$\{Y_i\}$, the per-sample false-negative loss at threshold $\tau$ is
$L_i(\tau) = \mathbf{1}\{Y_i = 1,\, S_i < \tau\}$, which is
non-decreasing and left-continuous in $\tau$ (equivalently,
non-increasing and right-continuous in $\lambda = -\tau$, the canonical
hypothesis of CRC Theorem~2.1). Conformal
Risk Control \cite{angelopoulos2024conformal} chooses
\begin{equation}
  \widehat{\tau} \;=\; \sup\!\Bigl\{\,\tau \;:\; \widehat{R}(\tau) + \tfrac{B}{n_\mathrm{cal\,haz}+1} \,\le\, \alpha\,\Bigr\},
  \label{eq:crc}
\end{equation}
where $\widehat{R}(\tau) = \tfrac{1}{n_\mathrm{cal\,haz}+1}\!\sum_{i\in\mathrm{cal\,haz}}\!\!L_i(\tau)$
is the empirical FNR on the $n_\mathrm{cal\,haz}$ calibration hazards,
and $B=1$ bounds the loss. Because the false-negative loss is zero on
benigns, the empirical mean is taken over hazards only; equivalently,
this is the class-conditional (Mondrian) instance of CRC
\cite{vovk2022algorithmic} run on the hazard subset, with the
guarantee $\mathbb{E}[\mathrm{FNR}] \le \alpha$ conditional on
$Y_{n+1} = 1$. Theorem 2.1 of
\citet{angelopoulos2024conformal} guarantees
$\mathbb{E}[L_{n+1}(\widehat{\tau})] \le \alpha$ when calibration and
test points are exchangeable. Under taxonomic-family holdout that
exchangeability is violated, and the bound picks up an additive total
variation term \cite{barber2023conformal}; we report the resulting full
right-hand side
\begin{equation}
  \mathbb{E}[\mathrm{FNR}] \;\le\; \alpha \;+\; \mathrm{TV}(\mathrm{cal},\mathrm{test}),
  \label{eq:bound}
\end{equation}
estimating the TV term by histogram TV between calibration and test
score distributions, a coarse but workable proxy.

\paragraph{Per-fold leak control.} Both $s_\mathrm{hom}$ and
$s_\mathrm{emb}$ are reference-set signals: a query's score depends on
which hazards sit in the reference. If we computed them once over the
full corpus and reused those values for every leave-one-family-out
fold, the test family's hazards would influence the score of their own
fold's queries through the reference set. We
recompute $s_\mathrm{hom}$ and $s_\mathrm{emb}$ inside each fold, using
\emph{only} the train-fold hazards as the reference set; the cached
global signals exist for inspection and are not consumed by the
evaluation loop. The LLM panel score does not use a hazard reference
set and is unchanged across folds.

\section{Experiments}
\label{sec:eval}

\subsection{Corpus, splits, hyperparameters}
\label{sec:eval:setup}

\paragraph{Corpus.} The hazard pool is UniProt KW-0800 (Toxin) restricted
to the reviewed Swiss-Prot subset \cite{uniprot2025}; this is a single
keyword query against the canonical public knowledgebase, and the same
keyword is the operational definition of ``toxin'' used by curators. The
benign pool is the reviewed Swiss-Prot subset minus KW-0800 and minus
KW-0843 (Virulence); the latter exclusion prevents virulence factors from
being mislabelled benign during evaluation. We sample $200$ hazards and
$400$ benigns ($n=600$ total, fixed seed) so that calibration sees a
$1{:}2$ hazard-to-benign ratio, a departure from the $\ll 1\%$
deployment ratio, chosen because pure deployment-ratio sampling
would leave so few hazards in the calibration set that the slack term
$1/(n_\mathrm{cal\,haz}+1)$ would dominate the bound. We address the
gap between this calibration ratio and deployment ratios in
\cref{sec:eval:budget}.

\paragraph{Splits.} Outer split: leave-one-taxonomic-family-out (LOTO),
a leave-one-group-out variant of the cross-validatory choice principle
\cite{stone1974crossvalidatory}, across the ten taxonomic families with
at least five hazards in the sample (Crotalinae, Hydrophiinae, Elapinae, Buthidae, Conus,
Theraphosidae, Sicariidae, Viperinae, Actiniidae, Lycosidae). For each
held-out family, the test set is every hazard from that family plus a
matched random sample of benigns at the corpus ratio. Inner split inside
the non-test pool: stratified $70/30$ train/calibration. Train fits the
aggregator weights; calibration chooses $\widehat{\tau}$. Within a fold,
the train, calibration, and test partitions are disjoint, and the
reference set for $s_\mathrm{hom}$ and $s_\mathrm{emb}$ is restricted to
train-fold hazards (\cref{sec:method:crc}).

\paragraph{Hyperparameters.} $k\!=\!5$ for the Jaccard signal,
$k_\mathrm{LLM}\!=\!2$ runs per (sample, model), temperature $0.7$,
$K\!\le\!8$ centroids, $\alpha=0.05$ unless stated otherwise. The
aggregator is logistic regression with sklearn's default $L_2$
regularisation ($C=1$) and the drop-and-refit non-negativity rule
(\cref{sec:method:agg}). All experiments use a fixed random seed (\cref{tab:hparams}).

\subsection{The bound holds on every fold}
\label{sec:eval:headline}

\begin{figure*}[!t]
  \centering
  \includegraphics[width=\textwidth]{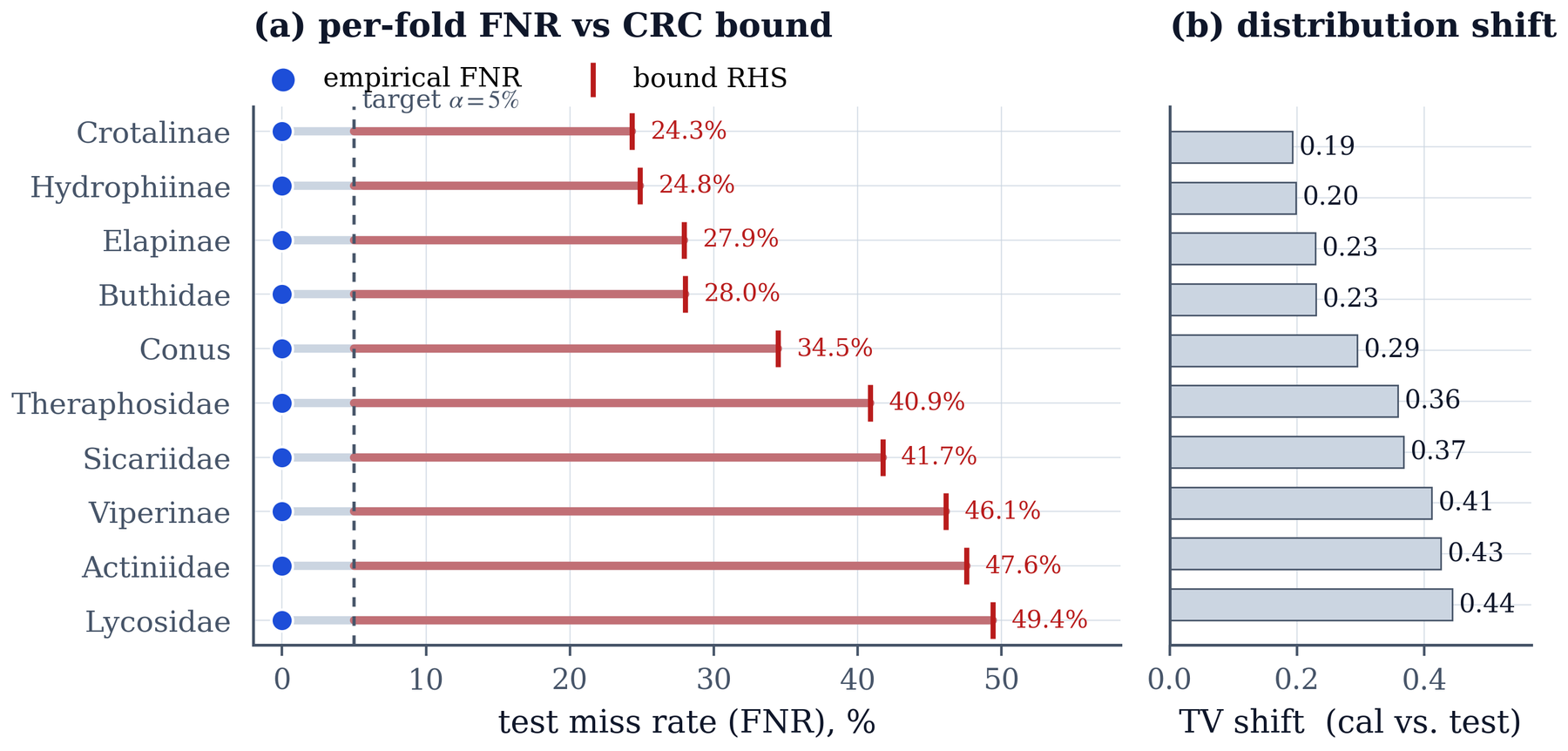}
  \caption{
    The CRC bound holds on every fold with $19$--$44$ percentage points
    of slack: empirical test miss rate is zero, while the certified
    ceiling $\alpha + \mathrm{TV}$ ranges from $24.3\%$ to $49.4\%$
    across ten LOTO folds at $\alpha=0.05$.
    (a) Per-fold view; the grey track runs from zero to the bound
    right-hand side, the red segment is the slack on top of $\alpha$,
    and the blue dot is the empirical test miss rate.
    (b) The TV proxy that drives the slack, $0.19$ (Crotalinae) to
    $0.44$ (Lycosidae). $n_\mathrm{cal\,haz}\in[51,58]$.
  }
  \label{fig:loto}
\end{figure*}

\Cref{fig:loto} shows the per-fold result at $\alpha\!=\!0.05$,
ordered by total-variation distance between the calibration and test
distributions of $S$; \cref{tab:perfold} lists the same per-fold
values, ordered by $n_\mathrm{test\,haz}$. Two findings:

\textbf{Empirical FNR is zero on every fold.} Across all ten folds the
calibrated screener misses zero hazards out of $5$ to $29$ test hazards
per family. Test FPR is also zero on nine of ten folds and $5\%$
(one of twenty test benigns) on Actiniidae.

\textbf{The bound is loose by design.} The right-hand side of
\cref{eq:bound} ranges from $24.3\%$ on Crotalinae, where the cal/test
TV is smallest, to $49.4\%$ on Lycosidae, where it is largest. The
bound is not vacuous: it certifies that the expected miss rate cannot
exceed roughly half on any fold under the observed distribution shift.

\begin{table*}[!t]
  \centering
  \small
  \begin{tabular}{lcccccccc}
    \toprule
    Family & $n_\mathrm{cal\,haz}$ & $n_\mathrm{test\,haz}$ & $n_\mathrm{test\,ben}$ & $\widehat{\tau}$ & TV proxy & test FNR & test FPR & bound RHS \\
    \midrule
    Crotalinae    & 51 & 29 & 58 & 0.598 & 0.193 & 0.000 & 0.000 & 0.243 \\
    Conus         & 53 & 22 & 44 & 0.719 & 0.295 & 0.000 & 0.000 & 0.345 \\
    Buthidae      & 53 & 21 & 42 & 0.675 & 0.230 & 0.000 & 0.000 & 0.280 \\
    Theraphosidae & 56 & 13 & 26 & 0.595 & 0.359 & 0.000 & 0.000 & 0.409 \\
    Hydrophiinae  & 56 & 11 & 22 & 0.622 & 0.198 & 0.000 & 0.000 & 0.248 \\
    Actiniidae    & 57 & 10 & 20 & 0.519 & 0.426 & 0.000 & 0.050 & 0.476 \\
    Elapinae      & 57 & 10 & 20 & 0.520 & 0.229 & 0.000 & 0.000 & 0.279 \\
    Viperinae     & 57 &  9 & 18 & 0.572 & 0.411 & 0.000 & 0.000 & 0.461 \\
    Sicariidae    & 57 &  8 & 16 & 0.659 & 0.367 & 0.000 & 0.000 & 0.417 \\
    Lycosidae     & 58 &  5 & 10 & 0.529 & 0.444 & 0.000 & 0.000 & 0.494 \\
    \bottomrule
  \end{tabular}
  \caption{Test FNR is zero on every fold and test FPR is zero on nine
    of ten folds (one flagged benign on Actiniidae) at $\alpha=0.05$,
    well inside the bound right-hand side $\alpha + \mathrm{TV}$;
    rows ordered by descending $n_\mathrm{test\,haz}$.}
  \label{tab:perfold}
\end{table*}

\subsection{Which signals carry the result}
\label{sec:eval:ablation}

\begin{figure*}[!t]
  \centering
  \includegraphics[width=0.92\textwidth]{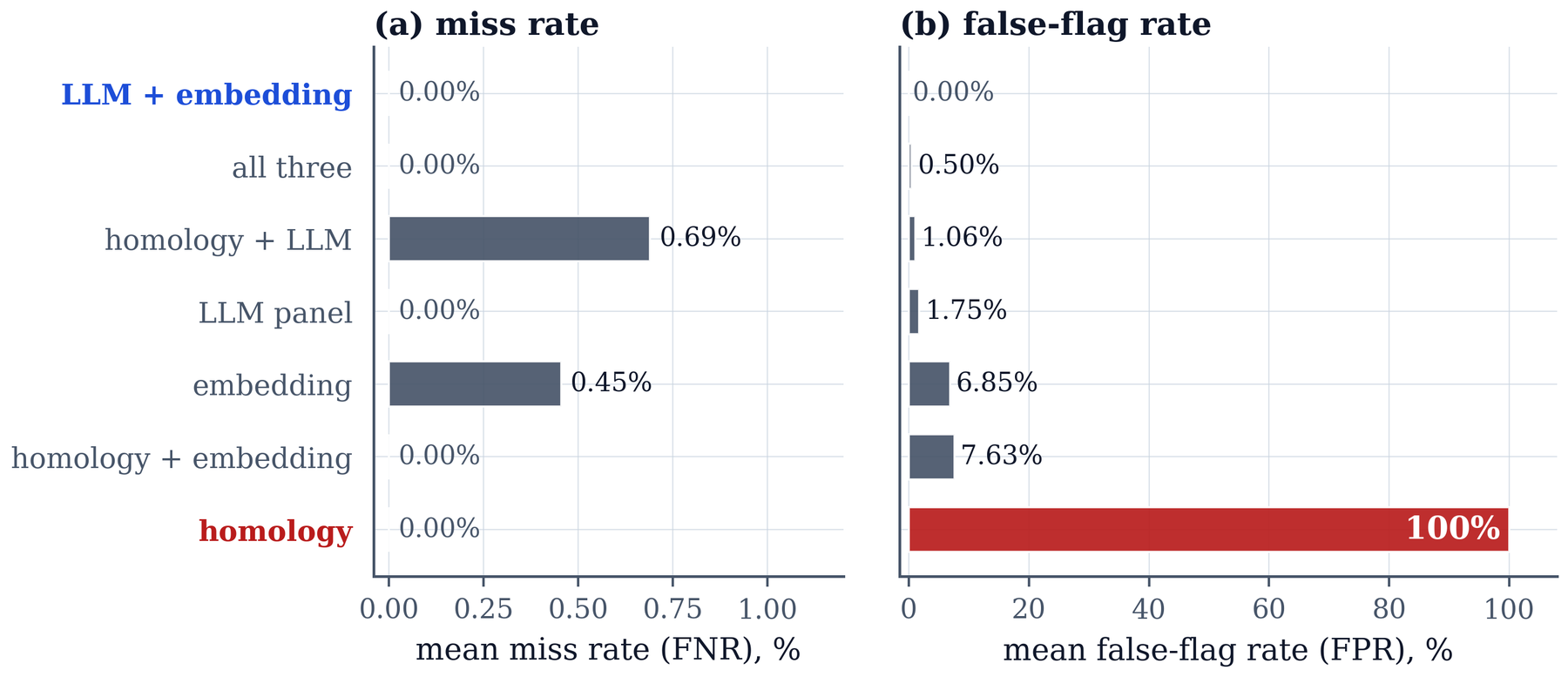}
  \caption{
    Of seven signal subsets at $\alpha=0.05$, only LLM+embedding
    achieves $0\%$ mean test FNR with $0\%$ mean test FPR; sequence
    homology alone fails at $100\%$ FPR; adding homology to LLM+embedding
    raises FPR to $0.5\%$ with no recall gain.
    Two combinations have non-zero mean FNR (embedding only: $0.45\%$;
    homology + LLM: $0.69\%$); their worst-fold FNRs ($4.55\%$ and
    $6.90\%$) are within the per-fold bound. Means across ten LOTO folds.
  }
  \label{fig:ablation}
\end{figure*}

To isolate which signals contribute to the headline result we re-run the
same per-fold protocol with each of the seven non-empty subsets of
$\{s_\mathrm{hom}, s_\mathrm{LLM}, s_\mathrm{emb}\}$ as input to the
aggregator and CRC. \Cref{fig:ablation} reports the mean test FNR and
mean test FPR across the ten folds at $\alpha=0.05$.

\textbf{Homology alone yields a $100\%$ false-flag rate.} Under
taxonomic-family holdout, the maximum $5$-mer Jaccard similarity
between any test-fold protein and any train-fold hazard is too low to
discriminate. Conformal Risk Control's coverage requirement then forces
$\widehat{\tau}$ down to (or below) the lowest calibration-hazard score,
which sits below the entire test-benign mass; the threshold becomes
``flag everything,'' and the resulting FPR is $1.0$ on every fold. This
is a structural failure mode of sequence-similarity screening when the
hazard at hand belongs to a family that the reference set has not seen,
not a tuning artefact.

\textbf{LLM panel and embedding each work alone, with caveats.} The
LLM panel alone achieves zero mean test FNR with $1.75\%$ mean FPR;
embedding alone achieves $0.45\%$ mean FNR (worst fold $4.55\%$) with
$6.85\%$ mean FPR. Either signal is sufficient on its own to avoid the
$100\%$ FPR pathology of homology, but neither alone hits the
$0\%/0\%$ profile.

\textbf{LLM panel + embedding is the operating point.} Composing the
two non-homology signals achieves $0\%$ mean test FNR and $0\%$ mean test
FPR, the headline result of \cref{fig:teaser}. \emph{Adding}
homology to this pair raises the mean FPR from $0\%$ to $0.5\%$ with no
recall gain. Homology is not merely unhelpful here; it is mildly
harmful as part of the ensemble, because
the train-fold-only reference set leaves a noisy near-uniform signal
that the aggregator weights into the score and CRC then has to budget
for.

\subsection{What $\alpha$ a given calibration set can certify}
\label{sec:eval:budget}

\begin{figure}[!htbp]
  \centering
  \includegraphics[width=\columnwidth]{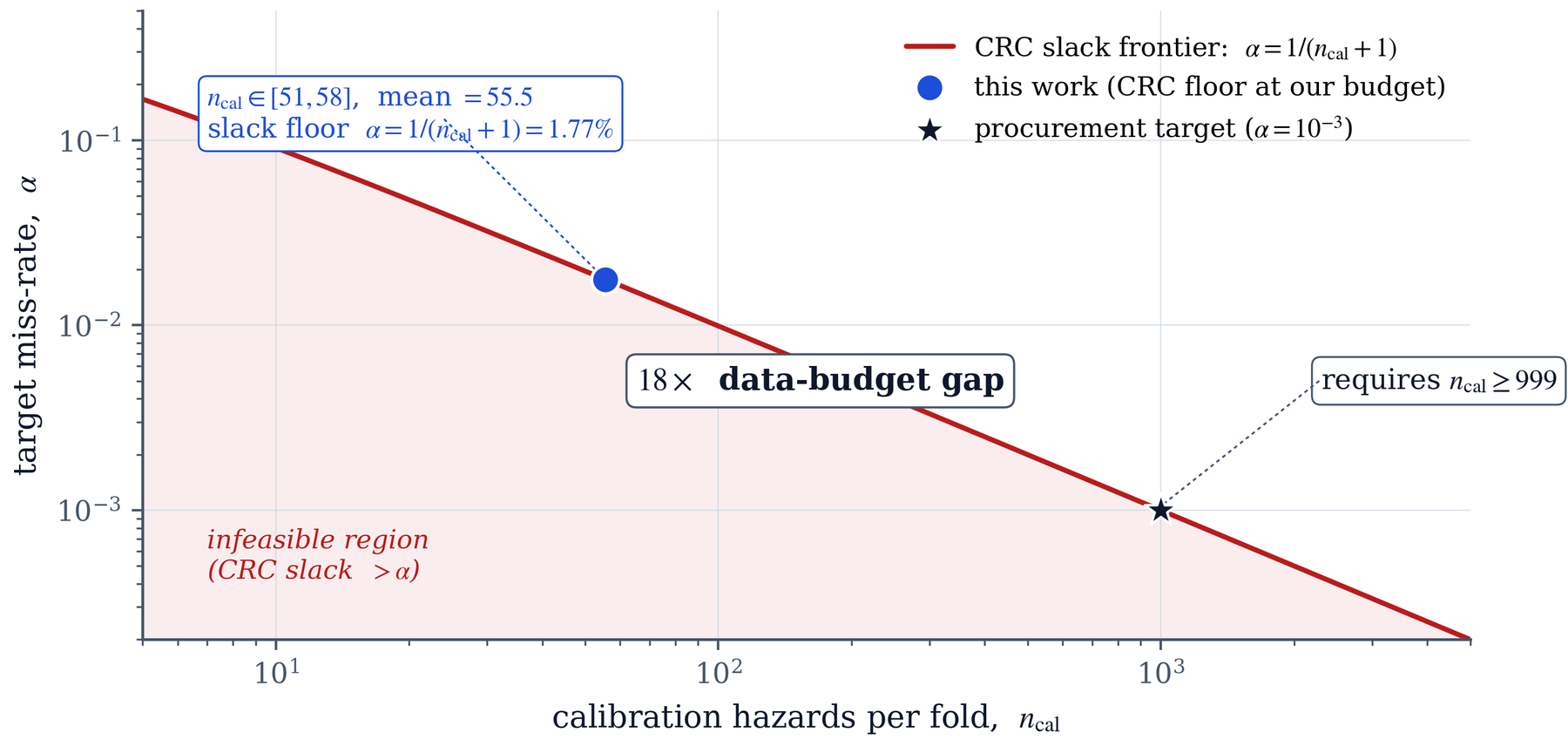}
  \caption{
    The CRC slack frontier $\alpha = 1/(n_\mathrm{cal}+1)$ caps the
    certifiable $\alpha$ at any calibration-set size; our $200$-hazard
    subsample ($n_\mathrm{cal}\!\approx\!55$, floor $1.77\%$) is
    $18\!\times$ below the procurement target $\alpha=10^{-3}$, but
    the full UniProt KW-0800 reviewed corpus has enough hazards to
    clear it. The shaded region is infeasible: any $\alpha$ below the
    frontier cannot be certified by Conformal Risk Control alone,
    regardless of model performance.
  }
  \label{fig:budget}
\end{figure}

The per-fold bound \cref{eq:bound} contains two slack terms beyond
$\alpha$. The first, $\mathrm{TV}(\mathrm{cal},\mathrm{test})$, is a
property of the splits: it shrinks if the calibration and test
distributions of $S$ become more similar. The second,
$1/(n_\mathrm{cal\,haz}+1)$, is a property of the calibration-set size
alone, and it sets a hard floor on the certifiable $\alpha$:
\begin{equation}
  \alpha \;<\; \frac{1}{n_\mathrm{cal\,haz}+1}
  \quad\Longrightarrow\quad \widehat{\tau}\!\to\!0,
  \label{eq:floor}
\end{equation}
because no $\tau$ can satisfy
$\widehat{R}(\tau) + 1/(n_\mathrm{cal\,haz}+1) \le \alpha$ when
$\widehat{R}\ge 0$ already exceeds $\alpha$. \Cref{fig:budget} traces this
frontier on a log-log axis.

On our subsample, $n_\mathrm{cal\,haz}\in[51,58]$ across folds, with
mean $55.5$, giving a slack floor of $\alpha\!\approx\!1.77\%$. Reaching
a stringency target of $\alpha=10^{-3}$ (an order-of-magnitude
deployment goal; we call this \emph{procurement-grade} below for
brevity) would require $n_\mathrm{cal\,haz}\!\ge\!999$, an
$18\!\times$ data-budget gap.
The full reviewed UniProt KW-0800 corpus contains roughly $6{,}000$
toxins; an evaluation at full scale would deliver
$n_\mathrm{cal\,haz}\!\sim\!1{,}800$ per fold, comfortably below the
procurement-grade floor.

\section{Discussion}
\label{sec:discussion}

\paragraph{Why composition works.} The three signals fail in different
directions: $s_\mathrm{hom}$ ignores function and is confounded by family-level
sequence drift; $s_\mathrm{LLM}$ ignores sequence and misreads
ambiguous annotations; $s_\mathrm{emb}$ misses fine-grained mechanism
and conflates topical similarity with functional similarity.
The aggregator's job is to recover from any single failure mode by
demanding agreement, and the monotone constraint forces this agreement
to be in the same direction for every signal. The empirical separation
in \cref{fig:ablation} between LLM-only ($1.75\%$ FPR) or embedding-only
($6.85\%$ FPR) and the LLM+embedding combination ($0\%$) is the
cooperative gain.

\paragraph{Why adding homology hurts.} Under taxonomic-family holdout
the per-fold homology score is essentially noise: it is rank-normalised
across the whole sample but computed against a reference set that
excludes the held-out family, so values in the test set are dominated
by random matches to unrelated train families. The aggregator weights
this noise with a small but non-negative coefficient, and Conformal
Risk Control then has to push $\widehat{\tau}$ slightly lower to absorb
the resulting calibration variance, which costs a fraction of a
percentage point in test FPR.

\subsection{Limitations}
\label{sec:limitations}

\emph{Sample size and seed variance.} We evaluate on a $200$-hazard
subsample with $n_\mathrm{cal\,haz}\!\approx\!55$ per fold and a
single random seed. The slack floor at this size caps the certifiable
$\alpha$ at $1.77\%$, far above procurement-grade $\alpha=10^{-3}$,
and per-fold empirical FNR of zero on $n_\mathrm{test\,haz}\in[5,29]$
carries wide Wilson confidence intervals. The three signals have
well-understood scaling behaviour, but multi-seed runs at full scale
are future work.

\emph{No comparison against fielded systems.} The IGSC and major
DNA-synthesis providers do not publish their screening procedures or
release reference sets \cite{diggans2019nextsteps}; we therefore
cannot directly compare CRC-Screen to a deployed baseline. Our homology-only condition is a stand-in for
the public alignment-search machinery, not for any specific commercial
implementation.

\emph{TV proxy versus true non-exchangeability bound.} Theorem~2 of
\citet{barber2023conformal} bounds the coverage gap by a weighted sum of
residual-swap TV terms; we substitute a histogram TV between
calibration and test score distributions. This is a coarse approximation
of the residual-swap quantity, not an upper bound, and the slack we
report could be larger or smaller than the exact bound depending on the
joint distribution.

\emph{Adversarial inputs are out of scope.} An adversary designing a
synthesis order to evade the screener would target the LLM-panel
(through annotation phrasing) or the embedding centroid (through
choice of organism / function description). Robustness against such
adversarial annotations is future work; the certified bound applies to
the joint cal/test distribution, not to a worst-case input.

\emph{Prompt-side LOTO leak.} The Variant~A prompt enumerates named
high-concern examples (e.g.~botulinum neurotoxin, ricin) that are
themselves UniProt KW-0800 entries. When one of these examples'
families is the held-out fold, the LLM panel has seen a name-level
description of the family in its prompt. We did not rotate the
example list per fold; doing so is a clean follow-up.

\paragraph{What changes in practice.} Two implications follow if these
results extrapolate. First, sequence-similarity-only screening is
insufficient as the sole defence under realistic distribution shift,
and the operational implication is that providers should compose at
least one annotation-derived signal (LLM panel, embedding distance, or
similar) into their screening stack. Second, the binding constraint on
certifiable miss-rates is the size of the labelled hazard pool used for
calibration, not the choice of model or aggregator: the algorithmic
tools to certify $\alpha=10^{-3}$ are available today, and the
investment required for procurement-grade screening is a larger,
better-curated calibration set.

\section{Conclusion}
\label{sec:conclusion}

Synthesis-order screening has been built as a sequence-matching
problem. Under taxonomic-family holdout that framing fails: sequence
similarity cannot deliver a certified miss rate without flagging
every benign, and the failure is not closeable by tuning. Recasting screening as a calibrated decision problem
closes it. Conformal Risk Control turns the operating threshold into
a data-driven calibration step with a certified miss-rate ceiling,
and the bound's two slack terms separate cleanly the part of the
problem that better algorithms can reduce from the part that only
more calibration data can. Three off-the-shelf signals clear the
bound at $\alpha=0.05$ today; the next factor of ten in certifiable
miss-rate comes from a larger, better-curated calibration set, not
from a better screener.

\section*{Impact Statement}

This work is defender-side: a synthesis provider screening incoming
orders for biosecurity-relevant proteins under a certified expected
miss rate. The system flags orders for human review and does not
generate, design, or modify biological sequences. The released code
contains no utility for sequence generation or pathogen-enhancement
information, no hazardous sequence data, and no operational guidance
for synthesis or expression.

The most plausible misuse pathway is an adversary with access to the
same public annotations and open-source tooling who scores their own
designs against the system to estimate evasion probability. The
threshold and aggregator weights shown in the paper are specific to
the demonstration fold and the $200$-hazard subsample, so they do not
transfer to any production screener trained on a larger hazard pool;
an adversary cannot read $\widehat{\tau}=0.60$ off this paper and
bypass a deployed system with it. The paper's most visible negative
finding, that sequence-similarity screening fails under taxonomic
holdout, is a property of how sequence similarity behaves under
distribution shift that has been documented in the open literature
\cite{puzis2020increased}; publishing it is consistent with
responsible-disclosure norms in the biosecurity community rather than
a new uplift.

The intended effect is defensive: to make certified-miss-rate
screening operationally available to providers, and to identify the
calibration set, not the algorithm, as the gap between current public
benchmarks and procurement-grade $\alpha=10^{-3}$ screening.

\bibliography{example_paper}

@inproceedings{angelopoulos2024conformal,
  title     = {Conformal Risk Control},
  author    = {Angelopoulos, Anastasios N. and Bates, Stephen and Fisch, Adam and Lei, Lihua and Schuster, Tal},
  booktitle = {The Twelfth International Conference on Learning Representations ({ICLR})},
  year      = {2024},
  url       = {https://openreview.net/forum?id=33XGfHLtZg}
}

@article{barber2023conformal,
  title     = {Conformal prediction beyond exchangeability},
  author    = {Barber, Rina Foygel and Cand{\`e}s, Emmanuel J. and Ramdas, Aaditya and Tibshirani, Ryan J.},
  journal   = {The Annals of Statistics},
  volume    = {51},
  number    = {2},
  pages     = {816--845},
  year      = {2023},
  doi       = {10.1214/23-AOS2276},
  publisher = {Institute of Mathematical Statistics}
}

@book{vovk2022algorithmic,
  title     = {Algorithmic Learning in a Random World},
  author    = {Vovk, Vladimir and Gammerman, Alexander and Shafer, Glenn},
  edition   = {2nd},
  publisher = {Springer Cham},
  year      = {2022},
  isbn      = {978-3-031-06648-1},
  doi       = {10.1007/978-3-031-06649-8}
}

@article{altschul1990basic,
  title   = {Basic Local Alignment Search Tool},
  author  = {Altschul, Stephen F. and Gish, Warren and Miller, Webb and Myers, Eugene W. and Lipman, David J.},
  journal = {Journal of Molecular Biology},
  volume  = {215},
  number  = {3},
  pages   = {403--410},
  year    = {1990},
  doi     = {10.1016/S0022-2836(05)80360-2}
}

@article{buchfink2021sensitive,
  title   = {Sensitive protein alignments at tree-of-life scale using {DIAMOND}},
  author  = {Buchfink, Benjamin and Reuter, Klaus and Drost, Hajk-Georg},
  journal = {Nature Methods},
  volume  = {18},
  number  = {4},
  pages   = {366--368},
  year    = {2021},
  doi     = {10.1038/s41592-021-01101-x}
}

@inproceedings{zheng2023judging,
  title     = {Judging {LLM}-as-a-Judge with {MT}-Bench and Chatbot Arena},
  author    = {Zheng, Lianmin and Chiang, Wei-Lin and Sheng, Ying and Zhuang, Siyuan and Wu, Zhanghao and Zhuang, Yonghao and Lin, Zi and Li, Zhuohan and Li, Dacheng and Xing, Eric P. and Zhang, Hao and Gonzalez, Joseph E. and Stoica, Ion},
  booktitle = {Advances in Neural Information Processing Systems 36 ({NeurIPS} 2023) Datasets and Benchmarks Track},
  year      = {2023},
  url       = {https://openreview.net/forum?id=uccHPGDlao}
}

@article{uniprot2025,
  title     = {{UniProt}: the Universal Protein Knowledgebase in 2025},
  author    = {{The UniProt Consortium}},
  journal   = {Nucleic Acids Research},
  volume    = {53},
  number    = {D1},
  pages     = {D609--D617},
  year      = {2025},
  doi       = {10.1093/nar/gkae1010},
  publisher = {Oxford University Press}
}

@article{pedregosa2011scikit,
  title   = {Scikit-learn: Machine Learning in {P}ython},
  author  = {Pedregosa, Fabian and Varoquaux, Ga{\"e}l and Gramfort, Alexandre and Michel, Vincent and Thirion, Bertrand and Grisel, Olivier and Blondel, Mathieu and Prettenhofer, Peter and Weiss, Ron and Dubourg, Vincent and Vanderplas, Jake and Passos, Alexandre and Cournapeau, David and Brucher, Matthieu and Perrot, Matthieu and Duchesnay, {\'E}douard},
  journal = {Journal of Machine Learning Research},
  volume  = {12},
  pages   = {2825--2830},
  year    = {2011},
  url     = {https://jmlr.org/papers/v12/pedregosa11a.html}
}

@article{puzis2020increased,
  title   = {Increased cyber-biosecurity for {DNA} synthesis},
  author  = {Puzis, Rami and Farbiash, Dor and Brodt, Oleg and Elovici, Yuval and Greenbaum, Dov},
  journal = {Nature Biotechnology},
  volume  = {38},
  number  = {12},
  pages   = {1379--1381},
  year    = {2020},
  doi     = {10.1038/s41587-020-00761-y}
}

@article{wittmann2025strengthening,
  title   = {Strengthening nucleic acid biosecurity screening against generative protein design tools},
  author  = {Wittmann, Bruce J. and Alexanian, Tessa and Bartling, Craig and Beal, Jacob and Clore, Adam and Diggans, James and Flyangolts, Kevin and Gemler, Bryan T. and Mitchell, Tom and Murphy, Steven T. and Wheeler, Nicole E. and Horvitz, Eric},
  journal = {Science},
  volume  = {390},
  number  = {6768},
  pages   = {82--87},
  year    = {2025},
  doi     = {10.1126/science.adu8578}
}

@article{bates2021distributionfree,
  author    = {Bates, Stephen and Angelopoulos, Anastasios and Lei, Lihua and Malik, Jitendra and Jordan, Michael},
  title     = {Distribution-free, Risk-controlling Prediction Sets},
  journal   = {Journal of the ACM},
  volume    = {68},
  number    = {6},
  pages     = {Article 43},
  year      = {2021},
  publisher = {Association for Computing Machinery},
  doi       = {10.1145/3478535}
}

@article{lei2018distributionfree,
  author    = {Lei, Jing and G'Sell, Max and Rinaldo, Alessandro and Tibshirani, Ryan J. and Wasserman, Larry},
  title     = {Distribution-Free Predictive Inference for Regression},
  journal   = {Journal of the American Statistical Association},
  volume    = {113},
  number    = {523},
  pages     = {1094--1111},
  year      = {2018},
  doi       = {10.1080/01621459.2017.1307116}
}

@inproceedings{tibshirani2019conformal,
  author    = {Tibshirani, Ryan J. and Foygel Barber, Rina and Cand{\`e}s, Emmanuel J. and Ramdas, Aaditya},
  title     = {Conformal Prediction Under Covariate Shift},
  booktitle = {Advances in Neural Information Processing Systems 32 (NeurIPS 2019)},
  year      = {2019}
}

@inproceedings{romano2019conformalized,
  author    = {Romano, Yaniv and Patterson, Evan and Cand{\`e}s, Emmanuel J.},
  title     = {Conformalized Quantile Regression},
  booktitle = {Advances in Neural Information Processing Systems 32 (NeurIPS 2019)},
  year      = {2019}
}

@article{sadinle2019least,
  author    = {Sadinle, Mauricio and Lei, Jing and Wasserman, Larry},
  title     = {Least Ambiguous Set-Valued Classifiers With Bounded Error Levels},
  journal   = {Journal of the American Statistical Association},
  volume    = {114},
  number    = {525},
  pages     = {223--234},
  year      = {2019},
  doi       = {10.1080/01621459.2017.1395341}
}

@article{cauchois2021knowing,
  author  = {Cauchois, Maxime and Gupta, Suyash and Duchi, John C.},
  title   = {Knowing What You Know: Valid and Validated Confidence Sets in Multiclass and Multilabel Prediction},
  journal = {Journal of Machine Learning Research},
  volume  = {22},
  number  = {81},
  pages   = {1--42},
  year    = {2021}
}

@article{angelopoulos2023gentle,
  author    = {Angelopoulos, Anastasios N. and Bates, Stephen},
  title     = {Conformal Prediction: A Gentle Introduction},
  journal   = {Foundations and Trends in Machine Learning},
  volume    = {16},
  number    = {4},
  pages     = {494--591},
  year      = {2023},
  doi       = {10.1561/2200000101}
}

@article{altschul1997gapped,
  author  = {Altschul, Stephen F. and Madden, Thomas L. and Sch{\"a}ffer, Alejandro A. and Zhang, Jinghui and Zhang, Zheng and Miller, Webb and Lipman, David J.},
  title   = {Gapped {BLAST} and {PSI-BLAST}: a new generation of protein database search programs},
  journal = {Nucleic Acids Research},
  volume  = {25},
  number  = {17},
  pages   = {3389--3402},
  year    = {1997},
  doi     = {10.1093/nar/25.17.3389}
}

@article{steinegger2017mmseqs2,
  author  = {Steinegger, Martin and S{\"o}ding, Johannes},
  title   = {{MMseqs2} enables sensitive protein sequence searching for the analysis of massive data sets},
  journal = {Nature Biotechnology},
  volume  = {35},
  number  = {11},
  pages   = {1026--1028},
  year    = {2017},
  doi     = {10.1038/nbt.3988}
}

@article{edgar2010search,
  author  = {Edgar, Robert C.},
  title   = {Search and clustering orders of magnitude faster than {BLAST}},
  journal = {Bioinformatics},
  volume  = {26},
  number  = {19},
  pages   = {2460--2461},
  year    = {2010},
  doi     = {10.1093/bioinformatics/btq461}
}

@article{eddy2011accelerated,
  author  = {Eddy, Sean R.},
  title   = {Accelerated Profile {HMM} Searches},
  journal = {PLOS Computational Biology},
  volume  = {7},
  number  = {10},
  pages   = {e1002195},
  year    = {2011},
  doi     = {10.1371/journal.pcbi.1002195}
}

@article{mistry2021pfam,
  author  = {Mistry, Jaina and Chuguransky, Sara and Williams, Lowri and Qureshi, Matloob and Salazar, Gustavo A. and Sonnhammer, Erik L. L. and Tosatto, Silvio C. E. and Paladin, Lisanna and Raj, Shriya and Richardson, Lorna J. and Finn, Robert D. and Bateman, Alex},
  title   = {{Pfam}: The protein families database in 2021},
  journal = {Nucleic Acids Research},
  volume  = {49},
  number  = {D1},
  pages   = {D412--D419},
  year    = {2021},
  doi     = {10.1093/nar/gkaa913}
}

@article{suzek2015uniref,
  author  = {Suzek, Baris E. and Wang, Yuqi and Huang, Hongzhan and McGarvey, Peter B. and Wu, Cathy H. and {The UniProt Consortium}},
  title   = {{UniRef} clusters: a comprehensive and scalable alternative for improving sequence similarity searches},
  journal = {Bioinformatics},
  volume  = {31},
  number  = {6},
  pages   = {926--932},
  year    = {2015},
  doi     = {10.1093/bioinformatics/btu739}
}

@article{camacho2009blastplus,
  author  = {Camacho, Christiam and Coulouris, George and Avagyan, Vahram and Ma, Ning and Papadopoulos, Jason and Bealer, Kevin and Madden, Thomas L.},
  title   = {{BLAST+}: architecture and applications},
  journal = {BMC Bioinformatics},
  volume  = {10},
  pages   = {421},
  year    = {2009},
  doi     = {10.1186/1471-2105-10-421}
}

@article{rives2021biological,
  author  = {Rives, Alexander and Meier, Joshua and Sercu, Tom and Goyal, Siddharth and Lin, Zeming and Liu, Jason and Guo, Demi and Ott, Myle and Zitnick, C. Lawrence and Ma, Jerry and Fergus, Rob},
  title   = {Biological structure and function emerge from scaling unsupervised learning to 250 million protein sequences},
  journal = {Proceedings of the National Academy of Sciences},
  volume  = {118},
  number  = {15},
  pages   = {e2016239118},
  year    = {2021},
  doi     = {10.1073/pnas.2016239118}
}

@article{lin2023evolutionary,
  author  = {Lin, Zeming and Akin, Halil and Rao, Roshan and Hie, Brian and Zhu, Zhongkai and Lu, Wenting and Smetanin, Nikita and Verkuil, Robert and Kabeli, Ori and Shmueli, Yaniv and dos Santos Costa, Allan and Fazel-Zarandi, Maryam and Sercu, Tom and Candido, Salvatore and Rives, Alexander},
  title   = {Evolutionary-scale prediction of atomic-level protein structure with a language model},
  journal = {Science},
  volume  = {379},
  number  = {6637},
  pages   = {1123--1130},
  year    = {2023},
  doi     = {10.1126/science.ade2574}
}

@article{elnaggar2022prottrans,
  author  = {Elnaggar, Ahmed and Heinzinger, Michael and Dallago, Christian and Rehawi, Ghalia and Wang, Yu and Jones, Llion and Gibbs, Tom and Feher, Tamas and Angerer, Christoph and Steinegger, Martin and Bhowmik, Debsindhu and Rost, Burkhard},
  title   = {{ProtTrans}: Toward Understanding the Language of Life Through Self-Supervised Learning},
  journal = {IEEE Transactions on Pattern Analysis and Machine Intelligence},
  volume  = {44},
  number  = {10},
  pages   = {7112--7127},
  year    = {2022},
  doi     = {10.1109/TPAMI.2021.3095381}
}

@article{madani2023progen,
  author  = {Madani, Ali and Krause, Ben and Greene, Eric R. and Subramanian, Subu and Mohr, Benjamin P. and Holton, James M. and Olmos, Jose Luis and Xiong, Caiming and Sun, Zachary Z. and Socher, Richard and Fraser, James S. and Naik, Nikhil},
  title   = {Large language models generate functional protein sequences across diverse families},
  journal = {Nature Biotechnology},
  volume  = {41},
  number  = {8},
  pages   = {1099--1106},
  year    = {2023},
  doi     = {10.1038/s41587-022-01618-2}
}

@article{diggans2019nextsteps,
  author  = {Diggans, James and Leproust, Emily},
  title   = {Next Steps for Access to Safe, Secure {DNA} Synthesis},
  journal = {Frontiers in Bioengineering and Biotechnology},
  volume  = {7},
  pages   = {86},
  year    = {2019},
  doi     = {10.3389/fbioe.2019.00086}
}

@techreport{mouton2024operational,
  author      = {Mouton, Christopher A. and Lucas, Caleb and Guest, Ella},
  title       = {The Operational Risks of {AI} in Large-Scale Biological Attacks: Results of a Red-Team Study},
  institution = {RAND Corporation},
  number      = {RR-A2977-2},
  year        = {2024},
  url         = {https://www.rand.org/pubs/research_reports/RRA2977-2.html}
}

@techreport{carter2015dna,
  author      = {Carter, Sarah R. and Friedman, Robert M.},
  title       = {{DNA} Synthesis and Biosecurity: Lessons Learned and Options for the Future},
  institution = {J. Craig Venter Institute},
  address     = {La Jolla, CA},
  year        = {2015},
  url         = {https://www.jcvi.org/research/dna-synthesis-and-biosecurity-lessons-learned-and-options-future}
}

@inproceedings{liu2023geval,
  author    = {Liu, Yang and Iter, Dan and Xu, Yichong and Wang, Shuohang and Xu, Ruochen and Zhu, Chenguang},
  title     = {{G-Eval}: {NLG} Evaluation using {GPT-4} with Better Human Alignment},
  booktitle = {Proceedings of the 2023 Conference on Empirical Methods in Natural Language Processing (EMNLP)},
  pages     = {2511--2522},
  year      = {2023},
  doi       = {10.18653/v1/2023.emnlp-main.153}
}

@inproceedings{dubois2023alpacafarm,
  author    = {Dubois, Yann and Li, Xuechen and Taori, Rohan and Zhang, Tianyi and Gulrajani, Ishaan and Ba, Jimmy and Guestrin, Carlos and Liang, Percy and Hashimoto, Tatsunori B.},
  title     = {{AlpacaFarm}: A Simulation Framework for Methods that Learn from Human Feedback},
  booktitle = {Advances in Neural Information Processing Systems 36 (NeurIPS 2023)},
  year      = {2023}
}

@inproceedings{guo2017calibration,
  author    = {Guo, Chuan and Pleiss, Geoff and Sun, Yu and Weinberger, Kilian Q.},
  title     = {On Calibration of Modern Neural Networks},
  booktitle = {Proceedings of the 34th International Conference on Machine Learning (ICML)},
  series    = {Proceedings of Machine Learning Research},
  volume    = {70},
  pages     = {1321--1330},
  year      = {2017}
}

@inproceedings{geifman2017selective,
  author    = {Geifman, Yonatan and El-Yaniv, Ran},
  title     = {Selective Classification for Deep Neural Networks},
  booktitle = {Advances in Neural Information Processing Systems 30 (NIPS 2017)},
  year      = {2017}
}

@incollection{platt1999probabilistic,
  author    = {Platt, John C.},
  title     = {Probabilistic Outputs for Support Vector Machines and Comparisons to Regularized Likelihood Methods},
  booktitle = {Advances in Large Margin Classifiers},
  editor    = {Smola, Alexander J. and Bartlett, Peter L. and Sch{\"o}lkopf, Bernhard and Schuurmans, Dale},
  publisher = {MIT Press},
  pages     = {61--74},
  year      = {1999}
}

@inproceedings{reimers2019sentencebert,
  author    = {Reimers, Nils and Gurevych, Iryna},
  title     = {{Sentence-BERT}: Sentence Embeddings using Siamese {BERT}-Networks},
  booktitle = {Proceedings of the 2019 Conference on Empirical Methods in Natural Language Processing and the 9th International Joint Conference on Natural Language Processing (EMNLP-IJCNLP)},
  pages     = {3982--3992},
  year      = {2019},
  doi       = {10.18653/v1/D19-1410}
}

@inproceedings{karpukhin2020dense,
  author    = {Karpukhin, Vladimir and Oguz, Barlas and Min, Sewon and Lewis, Patrick and Wu, Ledell and Edunov, Sergey and Chen, Danqi and Yih, Wen-tau},
  title     = {Dense Passage Retrieval for Open-Domain Question Answering},
  booktitle = {Proceedings of the 2020 Conference on Empirical Methods in Natural Language Processing (EMNLP)},
  pages     = {6769--6781},
  year      = {2020},
  doi       = {10.18653/v1/2020.emnlp-main.550}
}

@inproceedings{devlin2019bert,
  author    = {Devlin, Jacob and Chang, Ming-Wei and Lee, Kenton and Toutanova, Kristina},
  title     = {{BERT}: Pre-training of Deep Bidirectional Transformers for Language Understanding},
  booktitle = {Proceedings of the 2019 Conference of the North American Chapter of the Association for Computational Linguistics: Human Language Technologies, Volume 1 (Long and Short Papers)},
  pages     = {4171--4186},
  year      = {2019},
  doi       = {10.18653/v1/N19-1423}
}

@incollection{dietterich2000ensemble,
  author    = {Dietterich, Thomas G.},
  title     = {Ensemble Methods in Machine Learning},
  booktitle = {Multiple Classifier Systems},
  series    = {Lecture Notes in Computer Science},
  volume    = {1857},
  publisher = {Springer},
  address   = {Berlin, Heidelberg},
  pages     = {1--15},
  year      = {2000},
  doi       = {10.1007/3-540-45014-9_1}
}

@inproceedings{caruana2004ensemble,
  author    = {Caruana, Rich and Niculescu-Mizil, Alexandru and Crew, Geoff and Ksikes, Alex},
  title     = {Ensemble selection from libraries of models},
  booktitle = {Proceedings of the Twenty-First International Conference on Machine Learning (ICML)},
  year      = {2004},
  doi       = {10.1145/1015330.1015432}
}

@article{stone1974crossvalidatory,
  author  = {Stone, M.},
  title   = {Cross-Validatory Choice and Assessment of Statistical Predictions},
  journal = {Journal of the Royal Statistical Society: Series B (Methodological)},
  volume  = {36},
  number  = {2},
  pages   = {111--133},
  year    = {1974},
  doi     = {10.1111/j.2517-6161.1974.tb00994.x}
}

@article{jungo2005toxprot,
  author  = {Jungo, Florence and Bairoch, Amos},
  title   = {{Tox-Prot}, the toxin protein annotation program of the {Swiss-Prot} protein knowledgebase},
  journal = {Toxicon},
  volume  = {45},
  number  = {3},
  pages   = {293--301},
  year    = {2005},
  doi     = {10.1016/j.toxicon.2004.10.018}
}

@article{kaas2012conoserver,
  author  = {Kaas, Quentin and Yu, Rilei and Jin, Ai-Hua and Dutertre, S{\'e}bastien and Craik, David J.},
  title   = {{ConoServer}: updated content, knowledge, and discovery tools in the conopeptide database},
  journal = {Nucleic Acids Research},
  volume  = {40},
  number  = {D1},
  pages   = {D325--D330},
  year    = {2012},
  doi     = {10.1093/nar/gkr886}
}
\bibliographystyle{icml2026}

\newpage
\appendix
\onecolumn

\section{Panel aggregation}
\label{app:refusal}

For each (sample, model, run) we attempt one API call and parse a JSON
score in $[0,1]$. A call that errors out, returns malformed JSON, or
returns a score outside $[0,1]$ is filled with the neutral value
$0.5$. For each (sample, model) we take the median of the
$k_\mathrm{LLM}=2$ run scores; at $k=2$ this equals their mean. Let
$m_1,\dots,m_5$ be the five per-model scores for a sample, and let
$m_{(1)}\le m_{(2)}\le m_{(3)}\le m_{(4)}\le m_{(5)}$ their sorted order.
The panel score is the trimmed mean of the middle three:
\[
  s_\mathrm{LLM} \;=\; \tfrac{1}{3}\bigl(m_{(2)} + m_{(3)} + m_{(4)}\bigr).
\]
Trimming one extreme on each side limits the influence of a single
$0.5$ fallback (or any single outlier model) on the aggregate without
discarding any per-model evidence beyond the outermost.

\section{Hyperparameters}
\label{app:hparams}

\begin{table}[H]
  \centering
  \begin{tabular}{ll}
    \toprule
    Parameter & Value \\
    \midrule
    Corpus size $n$ & $600$ ($200$ hazards, $400$ benigns) \\
    $k$-mer length for Jaccard homology & $5$ \\
    LLM runs per (sample, model) & $2$ \\
    LLM temperature & $0.7$ \\
    Embedding model & text-embedding-3-large (OpenAI) \\
    Embedding normalisation & L\textsubscript{2} \\
    $K$-means cluster count & $\min(8,\ \lfloor n_\mathrm{train\,haz}/5\rfloor)$ \\
    Aggregator & logistic regression, $L_2$ with $C=1$ \\
    Non-negativity rule & drop most-negative coefficient and refit \\
    Train/calibration split & $70/30$, stratified by class label \\
    Histogram TV bin count & $50$, equal-width on $[0,1]$ \\
    Random seed & $42$ \\
    $\alpha$ targets & $\{0.10,\,0.05,\,0.01,\,0.005,\,0.001\}$ \\
    Default $\alpha$ & $0.05$ \\
    \bottomrule
  \end{tabular}
  \caption{Hyperparameters used throughout the evaluation. A single random
    seed is used; per-fold splits are deterministic.}
  \label{tab:hparams}
\end{table}

\section{Prompts for the LLM panel}
\label{app:prompt}

The headline results use a single prompt pair: the system message
(Variant~A, ``screening'') and the shared user message template. Both
are reproduced verbatim from the public code repository. Two
alternative system variants (``risk\_assessment'' with no rubric;
``minimal'' with task statement only) are present in the repository
for a sensitivity analysis but were not used to produce the headline
numbers.

\begin{promptbox}{System message (Variant A: screening)}
Task: You are a biosecurity screening analyst. Given a protein's public UniProt annotation, assign a hazard score in [0.0, 1.0] indicating how strongly the documented function warrants escalation to human expert review.

Score from the provided annotation text, drawing on your knowledge of protein families and regulatory classifications (e.g., CDC/USDA Select Agent lists) to interpret what that text means.

Rubric:

0.0 -- 0.2 \quad Benign\\
\hspace*{1em}Metabolic, structural, regulatory, or sensory function with no annotated toxicity, virulence, or pathogenicity.\\
\hspace*{1em}Examples: lactate dehydrogenase, collagen alpha chain, cytochrome c oxidase subunit, rhodopsin.

0.2 -- 0.5 \quad Low concern\\
\hspace*{1em}Host-defense, antimicrobial, or enzyme-inhibition activity not annotated as a toxin or virulence factor. Includes pathogen-origin proteins with narrow or food-safe hazard profiles.\\
\hspace*{1em}Examples: nisin (bacteriocin), non-pathogen phospholipase A2, pathogen-derived serine protease inhibitor with immune evasion annotation.

0.5 -- 0.8 \quad Elevated concern\\
\hspace*{1em}Explicitly annotated as a toxin, virulence factor, pore-forming protein, or viral attachment/fusion protein from any organism. Regulatory status is not required for this tier.\\
\hspace*{1em}Examples: Shiga toxin, streptolysin O, listeriolysin O, type III secretion effector, influenza hemagglutinin, SARS-CoV-2 spike protein.

0.8 -- 1.0 \quad High concern\\
\hspace*{1em}Matches a protein from the CDC/USDA Select Agent and Toxin list, or belongs to a protein class with documented biological weapons history.\\
\hspace*{1em}Examples: botulinum neurotoxin, ricin, abrin, anthrax protective antigen, Staphylococcal enterotoxin B, Clostridium perfringens epsilon toxin.

Output \textemdash{} return only a valid JSON object with exactly two fields:\\
\hspace*{1em}\{``score'': \textless float in [0.0, 1.0]\textgreater, ``rationale'': \textless string, at most 25 words, naming the annotation term(s) that determined the score\textgreater\}

Constraints:\\
1. Organism name may inform your score only when the function annotation corroborates a hazard concern. Do not assign a high score on organism identity alone.\\
2. Do not produce sequence data, synthesis routes, or pathogen enhancement information.\\
3. Always return a score. If the annotation is absent or wholly uninformative, return \{``score'': 0.5, ``rationale'': ``insufficient annotation''\}.
\end{promptbox}

The user message is rendered per sample by substituting four fields
from the protein's UniProt annotation. The function-text field is
truncated to 1{,}200 characters before substitution to respect model
context limits.

\begin{promptbox}{User message template}
Protein annotation:

\begin{tabular}{@{}l@{ : }l@{}}
Name & \{name\} \\
Organism & \{organism\} \\
Keywords & \{keywords\} \\
Function & \{function\_text\}
\end{tabular}
\end{promptbox}

\end{document}